# Stereocomplexation of Poly(Lactic Acid)s on Graphite Nanoplatelets: from Functionalized Nanoparticles to Self-Assembled Nanostructures.

Matteo Eleuteri [a], Mar Bernal [a], Marco Milanesio [b], Orietta Monticelli [c,*] Alberto Fina [a,*]

a- Dipartimento di Scienza Applicata e Tecnologia, Politecnico di Torino, Sede di Alessandria, Viale T. Michel 5, Alessandria, Italy
b- Dipartimento di Scienze e Innovazione Tecnologica, Università degli Studi del Piemonte Orientale, Viale T. Michel 11, Alessandria, Italy
c- Dipartimento di Chimica e Chimica Industriale, Università di Genova, Via Dodecaneso, 31, 16146 Genova, Italy

*corresponding authors: alberto.fina@polito.it; orietta.monticelli@unige.it

**Abstract**

The control of nanostructuration of graphene and graphene related materials (GRM) into self-assembled structures is strictly related to the nanoflakes chemical functionalization, which may be obtained via covalent grafting of non-covalent interactions, mostly exploiting π-stacking. As the non-covalent functionalization does not affect the $sp^2$ carbon structure, this is often exploited to preserve the thermal and electrical properties of the GRM and it is a well-known route to tailor the interaction between GRM and organic media. In this work, non-covalent functionalization of graphite nanoplatelets (GnP) was carried out with ad-hoc synthesized pyrene-terminated oligomers of polylactic acid (PLA), aiming at the modification of GnP nanopapers thermal properties. PLA was selected based on the possibility to self-assemble in crystalline domains via stereocomplexation of complementary poly(L-lactide) (PLLA) and poly(D-lactide) (PDLA) enantiomers. Pyrene-initiated PLLA and PDLA were indeed demonstrated to anchor to the GnP surface. Calorimetric and X-ray diffraction investigations highlighted the enantiomeric PLAs adsorbed on the surface of the nanoplatelets self-organize to produce highly crystalline stereocomplex domains. Most importantly, PLLA/PDLA stereocomplexation delivered a significantly higher efficiency in nanopapers heat transfer, in particular through the thickness of the nanopaper. This is explained by a thermal bridging effect of crystalline domains between overlapped GnP, promoting heat transfer across the nanoparticles contacts. This work demonstrates the possibility to enhance the physical properties of contacts within a percolating network of GRM via the self-assembly of macromolecules and opens a new way for the engineering of GRM-based nanostructures.

**Introduction**

Graphene and graphene-related materials (GRM) are one of the most intensively explored nanoparticles family in materials science owing to their superior properties.(Geim and Novoselov, 2007; Gómez-Navarro et al., 2007; Lee et al., 2008; Stoller et al., 2008; Balandin and Nika, 2012) In particular, graphene oxide (GO), reduced graphene oxide (RGO), multilayer graphene (MLG) and graphite nanoplatelets (GnP) attracted a wide research attention in the last decade.(Ferrari et al., 2015) Their peculiar mechanical, electrical and thermal properties make GRM ideal platforms for the construction of sophisticated nanostructured systems, which fabrication requires precise control of graphene chemistry, including chemical modifications with specific functional groups.(Rodriguez-Perez et al., 2013) In this field, the organic functionalization of GRM has led to the manufacturing of high-performance multifunctional graphene-based materials where covalent and non-covalent bonding provides bridges between adjacent layers.(Meng et al., 2013; Wan et al., 2018) However, the covalent attachment of any functional group affects the thermal and electronic properties of GRM because of the perturbation of their aromatic character. Therefore, the supramolecular functionalization is often preferable as it may preserve the structure and properties of non-oxidized GRM, while it simultaneously enables the attachment of specific organic moieties, through π-stacking or hydrophobic and electrostatic interactions.(Zhang et al., 2007; Choi et al., 2010; Liu et al., 2010; Cheng et al., 2012; Hsiao et al., 2013; Ji et al., 2015; Georgakilas et al., 2016) In particular, π-π stacking interactions between GRM and polyaromatic hydrocarbons, such as pyrene, perylene, hexabenzocoronene, have been widely investigated and validated for the preparation of GRM-based materials. (Björk et al., 2010; Georgakilas et al., 2012; Parviz et al., 2012; Hirsch et al., 2013; Wang et al., 2014) Non-covalent functionalization of GRM may indeed be exploited to tailor interfaces between nanoplatelets and the surrounding organic media or to control interfaces within GRM networks.

Compatibilization between GRM nanoplatelets and organic polymers have been widely studied to promote dispersion in the preparation of nanocomposites as well as to tailor their physical properties. (Kuilla et al., 2010; Potts et al., 2011; Mittal, 2014; Papageorgiou et al., 2017) Polyaromatic derivatives, mostly based on pyrene or perylene substituted with a dangling chain able to compatibilize towards the organic polymer have been largely studied, including the functionalization of GRM with polymers bearing pyrene groups.(Jingquan et al., 2010; Liang et al., 2012; Tong et al., 2013; Wang et al., 2015; Fina et al., 2018)

On the other hand, the modification of interfaces between GRM nanoplatelets within their percolating network is currently of high interest to enhance mechanical, electrical and thermal performance in GRM aerogels, foams and nanopapers. In particular, thermal properties of GRM nanopapers have attracted a significant research interest for the application as flexible heat spreaders. (Shen et al., 2014; Song et al., 2014; Xin et al., 2014; Renteria et al., 2015; Bernal et al., 2017; Bernal et al., 2018) However, the application of GRM based nanopapers is typically limited by their low mechanical resistance and brittleness. The incorporation of a limited amount of polymers into GRM nanopapers may enhance their toughness and deformability, while decreasing the heat transfer efficiency of the nanopaper, as polymers are well known for their typically low thermal conductivity. However, as the thermal conductivity of polymers is strongly dependent on chain orientation (Singh et al., 2014; Chen et al., 2016) and crystallinity (Choy et al., 1993; Ronca et al., 2017), the possibility to control local orientation and crystallization of macromolecules at the surface of GRM currently appears as a fascinating route to produce mechanically strong and high thermal conductivity nanostructures.

Several polymers have been shown to effectively nucleate on GRM(Ferreira et al., 2013; Manafi et al., 2014; Bidsorkhi et al., 2017; Colonna et al., 2017b), despite the self-organization of highly ordered crystalline domains onto the surface of GRM remains challenging. An interesting option to self-assemble crystalline domains is via stereocomplexation of complementary polymer enantiomers. For instance, polylactic acid (PLA) has two optically active enantiomers because of the chiral nature of the monomer: poly(L-lactide) (PLLA) and poly(D-lactide) (PDLA). PLLA and PDLA are able to co-crystallize into racemic stereocomplexes (SC), which have been studied mostly for their higher thermal stability and superior physical properties than the homocrystals of the homopolymers.(Tsuji et al., 1992; Tsuji, 2005; Hirata and Kimura, 2008) The incorporation of small amounts of GRM and functionalized GRM in PLLA/PDLA blends was also shown to promote the intermolecular coupling between the enantiomers, thus acting as nucleating agents for the stereocomplex.(Sun and He, 2012; Wu et al., 2013; Gardella et al., 2015; Xu et al., 2015; Xu et al., 2016; Yang et al., 2016; Zhang et al., 2017)

In this work, we report a facile approach to promote PLA SC formation onto the surface of GnP, exploiting the denser chain packing in the SC structure(Anderson and Hillmyer, 2006; Xu et al., 2006) to modify the physical properties of the contact between graphite nanoplatelets. Pyrene-end functionalized enantiomeric PLAs with low molecular weight were synthesized and used to supramolecularly modify GnP. Then, stereocomplexation of the PLLA and PDLA anchored to the GnP surface was carried out, to produce highly crystalline domains acting as junctions between nanoplatelets. Results obtained demonstrated the promotion of the intermolecular coupling between the enantiomeric PLAs adsorbed on the surface of the nanoplatelets, leading to the efficient stereocomplexation of the systems. More importantly, the presence of highly crystalline domains from PLLA/PDLA stereocomplexation delivered nanopapers with a higher through-plane efficiency in heat transfer, acting as efficient thermal bridges between overlapped GnP.

**Experimental Section**

**Materials**

L-lactide and D-lactide (purity ≥ 98 %) were purchased from Sigma-Aldrich. Before polymerization, both monomers, L-lactide and D-lactide, were purified by three successive re-crystallizations from 100% (w/v) solution in anhydrous toluene and dried under vacuum at room temperature. 1-Pyrenemethanol (Pyr-OH) (purity 98 %) and stannous octanoate, tin(II) 2-ethylhexanoate (Sn(Oct)$_2$, purity ~ 95%) were purchased from Sigma-Aldrich and used as received. All solvents, toluene (anhydrous, purity 99.8 %), chloroform (purity ≥ 99.5 %), methanol (purity ≥ 99.8 %), N-N-dimethylformamide (DMF) (anhydrous, purity 99.8 %) and dimethyl ether (DME) (purity ≥ 99.0 %) were purchased from Sigma Aldrich and used without further purifications. Graphite nanoplatelets (GnP) were synthesized by Avanzare (Navarrete, La Rioja, Spain) according to a previously reported procedure.(Colonna et al., 2017a) In brief, overoxidized-intercalated graphite was prepared starting from natural graphite, followed by rapid thermal expansion of at 1000°C, to produce a worm-like solid, and then mechanically milled to obtain GnP.

**Synthesis of pyrene-end functionalized PLLA and PDLA (Figure 1a)**

Pyrene-end functionalized poly(L-lactic acid) (Pyr-L) and pyrene-end functionalized poly(D-lactic acid) (Pyr-D) were synthesized by the ring-opening polymerization (ROP) of monomers, L-lactide and D-lactide, initiated with Pyr-OH and catalyzed by Sn(Oct)$_2$ in bulk at 140°C, as previously reported.(Eleuteri et al., 2018) In detail, L-lactide or D-lactide (5.5 g, 38 mmol) were charged under argon flow into the reactor (*i.e.*, a 50-ml two-neck round-bottomed flask equipped with a magnetic stirrer). Pyr-OH (220 mg, 0.94 mmol) was introduced under argon flow and the flask was evacuated for 15 min and purged with argon; these vacuum/argon cycles were repeated three times in order to thoroughly dry the reactants. The monomer-to-Pyr-OH molar ratio was calculated to obtain an average theoretical molecular weight ($M_{nth}$) for Pyr- L or Pyr- D of 7000 g/mol. Then, the reactor vessel was immersed in an oil bath at 140 °C under stirring. Once the reactants were completely melted and homogenized, a freshly prepared solution of Sn(Oct)$_2$ in toluene (271 µL, [lactide]/[Sn(Oct)$_2$] = 10$^3$) was added under argon and the reaction was allowed to proceed under inert atmosphere for 24h. After cooling to room temperature, the reaction was quenched in an ice bath and the crude products were dissolved in chloroform and poured into an excess of cold methanol (2°C). The solid residue was filtered and dried in vacuum at 40 °C. Then, Pyr-L was dissolved in DMF (15 mg mL$^{-1}$) at 65 °C for 15 min. The solvent was allowed to evaporate at 25°C for 48 hours and the residual solvent was removed by drying at 60 °C for 3 hours.

**Preparation of Pyr-L/Pyr-D stereocomplex (Figure 1b)**

Equivalent amounts of Pyr-L and Pyr-D were separately dissolved in DMF (15 mg mL$^{-1}$) at 65 °C. The solutions were then mixed and stirred at 65 °C for 15 min. The mixed solution was casted on a Petri dish. The solvent was allowed to evaporate at 25°C for 48 hours and the residual solvent was removed by drying at 60 °C for 3 hours. The above-described blends were denoted as Pyr-L/Pyr-D SC. It is worth mentioning that both parallel and antiparallel stereocomplex structures may be obtained from the blend of Pyr-L and Pyr-D, as previously reported for similar pyrene-terminated PLA oligomers.(Danko et al., 2018) However, the two structures have very limited differences in cell dimensions(Brizzolara et al., 1996) and interaction energies, so that the two cannot be distinguished by XRD data and thermal analyses and, most importantly, are expected to deliver very similar material properties.

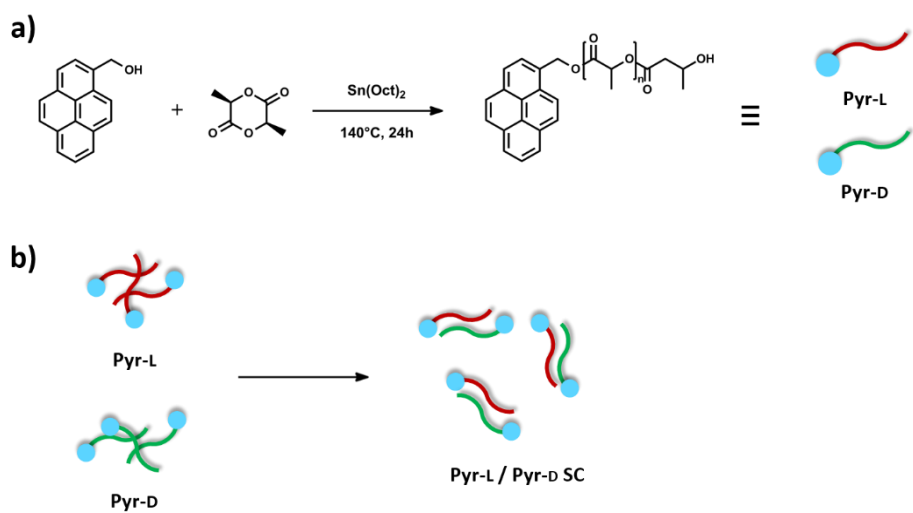

**Figure 1.** (a) Synthesis of Pyr-L and Pyr-D. (b) Schematic illustration of stereocomplex formation from equivalent amounts of Pyr-L and Pyr-D.

**Preparation of functionalized GnP Pyr-L (Figure 2a)**

100 mg GnP were added to a 100 mL solution of Pyr-L in DMF (0.5 mg·mL$^{-1}$) and sonicated in pulsed mode (30 s ON and 30 s OFF) for 15 min with power set at 30 % of the full output power (750 W) by using an ultrasonication probe (Sonics Vibracell VCX-750, Sonics & Materials Inc.) with a 13 mm diameter Ti-alloy tip. After that, the suspension was left to stand for 24 hours to ensure the adsorption of the stereoisomer on the basal plane of GnP. Then, the suspension was filtered through a Nylon Supported membrane (0.45 μm nominal pore size, diameter 47 mm, Whatman) and the filtrate analysed by UV-Vis. The filtered cake was re-dispersed in DMF (50 mL), sonicated in an ultrasonication bath for 10 min, filtered and the filtrate analyzed by UV-Vis. The dispersion-filtration cycle was repeated a second time to ensure the completely disappearance of the Pyr-L on the filtrate. Finally, the functionalized GnP were washed with methanol (50 mL) and diethyl ether (50 mL) and dried at 60 °C for 24 hours.

**Preparation of functionalized GnP Pyr-L/Pyr-D SC (Figure 2b and c)**

Functionalized GnP Pyr-L/Pyr-D SC were prepared following two different methods. In the first method (Figure 2b), 100 mg GnP was added to a 100 mL DMF solution containing equivalent masses of Pyr-L (25 mg) and Pyr-D (25 mg). The procedure previously described for sonication, filtration and purification of the functionalized GnP Pyr-L was followed and the modified GnP obtained by this method are coded as GnP Pyr-L/Pyr-D SC_A. In the second method (Figure 2c), two suspensions of GnP Pyr-L and GnP Pyr-D (25 mg of enantiomer and 50 mg GnP in 50 mL DMF) prepared as described previously for GnP Pyr-L were mixed together to obtain a final solution with the same concentrations of enantiomers and GnP as in GnP Pyr-L/Pyr-D SC_A. The procedure previously described for sonication, filtration and purification was followed and the modified GnP obtained by this method are coded as GnP Pyr-L/Pyr-D SC_B.

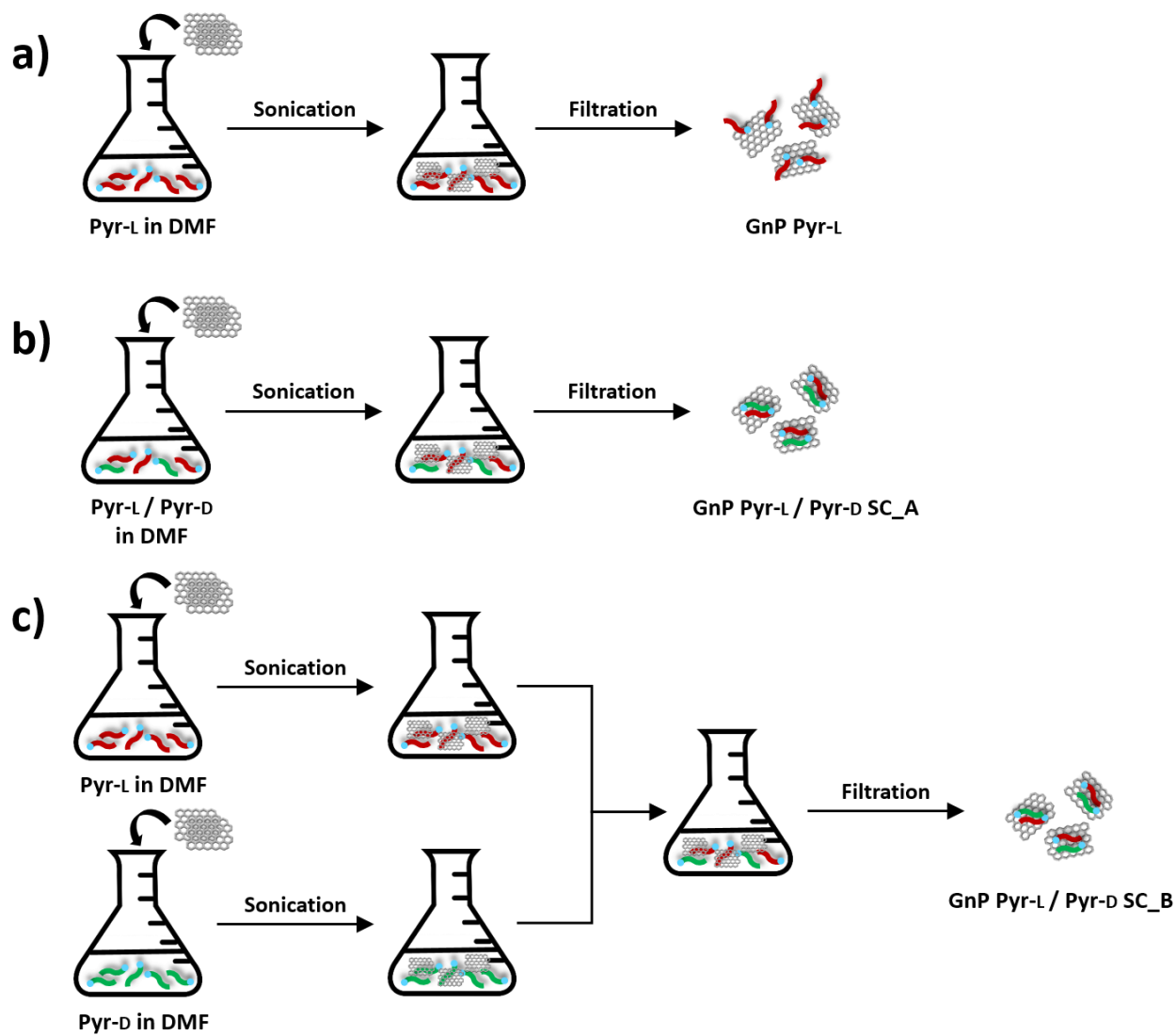

**Figure 2.** Preparation of functionalized GnP: a) GnP Pyr-L, b) GnP Pyr-L/Pyr-D SC_A and c) GnP Pyr-L/Pyr-D SC_B.

**Preparation of GnP and functionalizedGnP nanopapers**

GnP nanopapers were prepared by suspension of GnP in DMF at a concentration of 0.2 mg mL$^{-1}$ and the solutions were sonicated as described previously for the preparation of functionalized GnP. Then, GnP nanopapers were prepared by vacuum filtration, dried and mechanically pressed, following the procedure previously reported.(Bernal et al.; Bernal et al., 2017)

GnP nanopapers containing the stereocomplexes were prepared following the method described previously for GnP Pyr-L/Pyr-D SC_A. The total concentrations of Pyr-L and Pyr-D in DMF were 0.1 mg mL$^{-1}$, while the final concentration of GnP in the solutions was 0.2 mg mL$^{-1}$. For comparison purposes, two additional nanopapers were prepared for the formation of the stereocomplex, increasing the concentrations of the stereoisomers to 0.2 and 0.4 mg mL$^{-1}$, denoted as GnP Pyr-L/Pyr-D SC_A′ and GnP Pyr-L/Pyr-D SC_A″, respectively.

**Characterization methods**

All $^1$H Nuclear Magnetic Resonance (NMR) spectra were recorded on a NMR Varian Mercury Plus 300 MHz. Samples were dissolved in deuterated chloroform (CDCl$_3$) with TMS as internal reference (chemical shifts δ in ppm).

Differential scanning calorimetry (DSC) measurements were performed on a DSC Q20 (TA Instruments , USA). Approximately 5 mg of sample were placed in the aluminum pans. The samples were heated from 25 °C to 250 °C at a heating rate of 10 °C min$^{-1}$ and kept for 1 min to erase the thermal history. Afterwards, the specimens were cooled down to 25 °C, and finally reheated to 250 °C at 10 °C min$^{-1}$ to evaluate the crystallization and melting behaviour of the samples. For functionalized GnP, the reference aluminum pan was filled with pristine GnP to facilitate the observation of the thermal transitions arising from the Pyr-L and Pyr-L/Pyr-D SC.

Powder X-ray diffraction (XRD) measurements for the stereoisomers, the SC crystallites and functionalized GnP were carried out using an X-ray diffractometer (PANalytical X'Pert Pro MPD, Philips PW3040/60) with a Cu Kα radiation source with a wavelength of 1.542 Å. The measurements were operated at 40 kV and 40 mA with scan angles from 5° to 25° at a scan step of 0.026°. XRD measurements on nanopapers were carried out also with a 2D detector, exploiting a a Gemini R Ultra diffractometer. All data were collected using Cu Kα radiation. Data collection and reduction was carried out with CrysAlisPro software, version 1.171.35.11. (Agilent Technologies UK Ltd. Oxford, UK). 2D images were collected with a time from 10 to 30 seconds depending on the 2θ. The 2D data were then reduced to intensity vs. 2θ profiles by the same software, to investigate preferential orientations in the nanopapers and well as to obtain averaged XRD profiles over the whole possible orientations.

Thermal gravimetrical analysis (TGA) were performed using a TGA Discovery (TA Instruments, USA) under nitrogen atmosphere from 50 °C to 800 °C at 10 °C min$^{-1}$. The thermal degradation temperatures were defined as the temperatures corresponding to the maximum DTG peaks, obtained from the first derivative curve of the TGA thermogram.

Fourier-transform infrared (FT-IR) spectra were recorded on a Perkin Elmer Perkin Elmer Frontier spectrometer (Waltham, MA, USA) in the range of 400 – 4000 cm$^{-1}$ with 16 scans at a resolution of 4 cm$^{-1}$.

X-ray photoelectron spectroscopy (XPS) were performed on a VersaProbe5000 Physical Electronics X-ray photoelectron spectrometer with a monochromatic Al source and a hemispherical analyzer. Survey scans and high-resolution spectra were recorded with a spot size of 100 μm. The samples were prepared by depositing the GnP powders onto adhesive tape and keeping the samples under vacuum for 15 h prior to the measurement. A Shirley background function was employed to remove the background of the spectra.

The morphology of the graphene papers was characterized by a high resolution Field Emission Scanning Electron Microscope (FESEM, ZEISS MERLIN 4248).

The density ($\rho$) of the nanopapers was calculated according to the formula $\rho = m/V$, where $m$ is the mass of the nanopaper, weighed at room temperature using the TGA microbalance (Sensitivity: < 0.1 μg) and $V$ is calculated from a well-defined disk film using the average thicknesses measured by FESEM.

The in-plane thermal diffusivity ($\alpha_{//}$) and cross-plane diffusivity ($\alpha_{\perp}$) were measured using the xenon light flash technique (LFT) (Netzsch LFA 467 *Hyperflash*). The samples were cut in disks of 23 mm and the measurement of the $\alpha_{//}$ was carried out in a special in-plane sample holder while the $\alpha_{\perp}$ was measured in the standard cross-plane configuration. Each sample was measured five times at 25 °C.

**Results and Discussion**

**Synthesis of Pyr-L and Pyr-D by ROP and formation of Pyr-L/Pyr-D SC**

The Pyr-L and Pyr-D synthesized by the ring-opening polymerization of L-lactide and D-lactide using Pyr-OH as initiator have been first characterized for their molar masses. The number average molecular weight ($M_{nNMR}$) of Pyr-L and Pyr-D was estimated based on the $^1$H-NMR spectra (Figure S1), using the integral area of the methine protons signals in the PLA chain and next to the terminal hydroxyl group, at $\delta = 5.16$ ppm and $\delta = 4.35$ ppm, respectively. The $M_{nNMR}$ is *ca.* 7000 g mol$^{-1}$, which is in accordance to the theoretical number average molecular weight ($M_{nth}$) calculated from the monomer-to-Pyr-OH molar ratio, thus confirming the controlled polymerization reaction.

Furthermore, the characteristic absorption bands of pyrene (Figure S2) confirm the presence of the cromophore group in the enantiomers.

Crystallization of as-synthesized Pyr-L or Pyr-L/Pyr-D SC was obtained after dissolving the product in N,N-dimethylformamide (DMF) and left to crystallize. DMF was chosen in this study, allowing for the formation of stereocomplex crystallites and providing sufficient affinity for the subsequent dispersion of graphene nanoplatelets, as the surface tension of DMF (37.1 mJ m$^{-2}$) is similar to that of graphene.(Hernandez et al., 2008; Coleman et al., 2011) Given the influence of the solvent on the polymorphic crystalline structure of poly(L-lactide)(Marubayashi et al., 2012; 2013) and stereocomplex crystallization of PLA(Tsuji and Yamamoto, 2011; Yang et al., 2017) have been previously reported, for comparison purposes, Pyr-L or Pyr-L/Pyr-D SC crystallizations were also performed in chloroform, being the most common solvent for PLA (result comparison in Supporting Information).

The structures of the Pyr-L and Pyr-L/Pyr-D SC after crystallization from DMF were corroborated by FTIR spectroscopy (Figure 3). The $v$(C=O) band, which is sensitive to the morphology and the conformation of PLA,(Kister et al., 1998) changed after the stereocomplexation, showing a downshift of the signal which is related to the weak hydrogen bond formation between the CH$_3$ groups and the C=O group in the Pyr-L/ Pyr-D SC.(Kister et al., 1998; Zhang et al., 2005; Gonçalves et al., 2010) Furthermore, the CH$_3$ and CH bending region, between 1250 and 1400 cm$^{-1}$, showed asymmetric and broad bands for Pyr-L compared to Pyr-L/Pyr-D SC, being the δ(CH) band high-frequency shift for the stereocomplex (~ 10 cm$^{-1}$), as previously observed by Kister *et al.*(Kister et al., 1998) The band at 920 cm$^{-1}$, observed in Pyr-L and assigned to the α-helix of an enantiomeric PLA, disappeared in and Pyr-L/Pyr-D SC and a new band appeared at 909 cm$^{-1}$ , corresponding to β-helix, characteristic of the stereocomplex, accordingly to previous report by Michalski *et al.*(Michalski et al., 2018) Finally, after stereocomplexation the $v$(C-COO) stretching mode at 872 cm$^{-1}$ is shifted at higher frequency and appears less asymmetric compared to the corresponding band in Pyr-L spectrum, thus further confirming the successful formation of the stereocomplex crystallites.

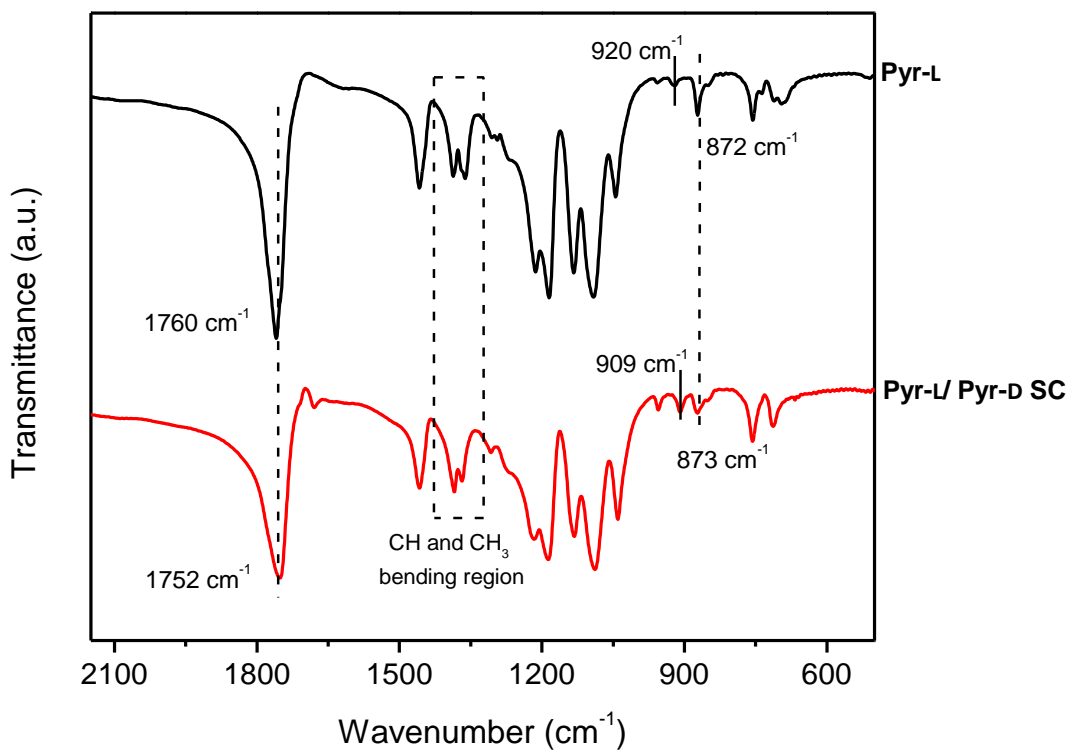

**Figure 3.** FTIR spectra (500 – 2250 cm$^{-1}$ region) of Pyr-L (black) and Pyr-L/Pyr-D SC (red) after crystallization in DMF.

The stereocomplexation of pyrene-based PLA systems were analyzed by DSC (Figure 4a) and XRD (Figure 4b). The stereoisomers, Pyr-L and Pyr-D, were confirmed to crystallize in the stereocomplex form as demonstrated by the increase of the melting temperature ($T_m$), from 152.2 °C for Pyr-L to 207.5 °C for Pyr-L/Pyr-D SC. Moreover, the formation of the Pyr-L/Pyr-D SC leads to both higher enthalpy of melt crystallization ($\Delta H_m$) and high $T_m$ in DMF compared to the stereocomplex formed in chloroform (see Figure S3 and Table S1), evidencing for the crucial role of the solvent in promoting stereocomplexation. Interestingly, no crystallization peak ($T_{cc}$) at around 100 °C was observed during the formation of Pyr-L/Pyr-D SC in DMF, which is ascribed to the crystallization during the previous cooling scan.(Bao et al., 2016) Hence, the π-π interactions between pyrene groups in low molecular weight Pyr-L and Pyr-D did not hinder the stereocomplex crystallization,(de Arenaza et al., 2013) as exclusive stereocomplex crystallites are formed in DMF.(Yang et al., 2017) The effect of DMF on the crystallization of Pyr-L and the formation of SC crystallites in Pyr-L/Pyr-D SC were also confirmed by XRD (Figure 4b). In fact, Pyr-L shows the typical diffraction peaks of the stereoisomer at 2θ 14.8°, 16.7°, 19.1° and 22.4°, corresponding to the (010), (110)/(200), (203) and (210) planes. Additionally, the crystalline complex ε-form, induced by the formation of complexes in specific solvents with five-membered ring structure such as DMF,(Marubayashi et al., 2012; 2013) is evidenced by the peak observed at 2θ ≈ 12.4°. Pyr-L/Pyr-D SC formed in DMF exhibits only the diffraction peaks of SC crystallites at 12.1°, 21.1° and 24.0°, thus indicating the exclusive formation of SC crystallites, due to the low vapor pressure of DMF and the dissimilar solubility parameter between solvent and polymer.(Tsuji and Yamamoto, 2011; Yang et al., 2017)

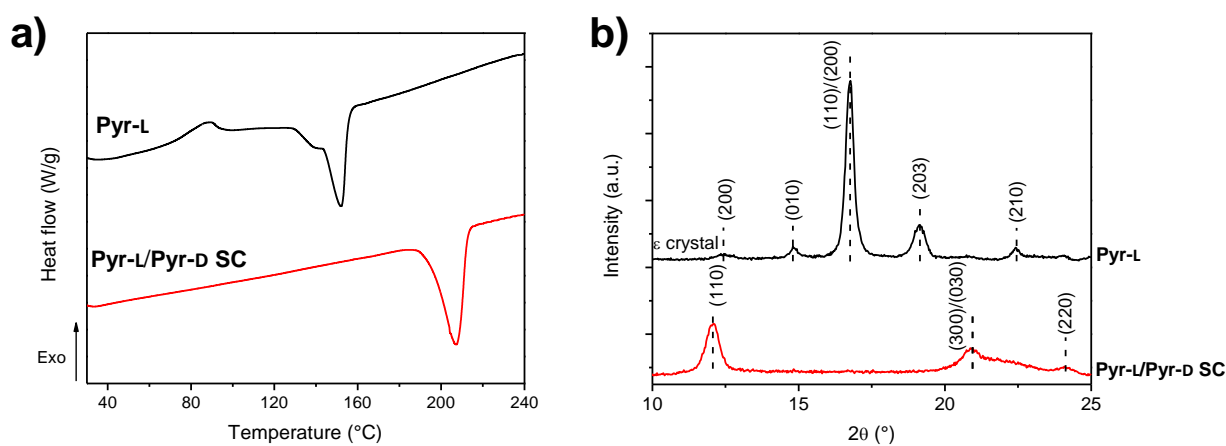

**Figure 4.** a) DSC thermograms of Pyr-L and Pyr-L/Pyr-D SC recorded during the second heating (10 °C min$^{-1}$). b) Powder XRD patterns of Pyr-L and Pyr-L/Pyr-D SC.

**Supramolecular functionalization of GnP with pyrene-end functionalized PLAs**

The formation of the stereocomplex in the presence of GnP was investigated using two different approaches: i) sonicating the nanoplatelets in a solution containing both Pyr-L and Pyr-D (Figure 2b) and ii) sonicating separately the GnP with each of the stereoisomers, Pyr-L or Pyr-D, before mixing them to form the stereocomplex (Figure 2c). After that, the products were washed thoroughly to remove any DMF-soluble oligomers unbound to GnP (See Figure S4 in the Supporting Information).

The modification of GnP was investigated by XPS (Figure S5), which provides information about the elemental composition and chemical binding states of the samples. Pyr-L and Pyr-L/Pyr-D SC were firt analysed, showing signals at ~ 284.9 eV attributed to aliphatic carbons (-CH$_2$, -CH$_3$), at 285.5 eV (C-OH), the C-O-C peak at 286.8 eV and the 288.9 eV binding energy of the -C=O. The deconvolution of the $C_{1s}$ spectra of the functionalized GnP are dominated by the peak at 284.4 eV, related to the $sp^2$ C-C bonds and thus the differentiation of the additional aliphatic carbons of the polymer backbone was difficult, suggesting for a limited fraction of PyrL or PyrD onto GnP. However, the $O_{1s}$ spectra (Figure S5b) showed the three bands typical of the composition of PLA (O-C=O at 531.5 eV, C-O-C at 533.0 eV and C-OH at 534.2 eV) reflected in GnP Pyr-L and GnP Pyr-L/Pyr-D SC, evidencing significant differences compared to pristine GnP and directly evidencing for grafted PLA chains. The fitting of XPS spectra, with additional comments and atomic ratios for each oxygen functional group are reported in Supporting Info, Figure S5 and Table S2.

The functionalization of GnP by the stereoisomers and the formation of the stereocomplex on the basal plane of GnP was also corroborated by thermal analysis (Figure 5). Pristine GnP showed a 3.0 % weight loss until 600°C under $N_2$ atmosphere, because of the decomposition of organic impurities and or elimination of the few oxidized groups obtained during the synthesis process of such nanoplatelets.(Colonna et al., 2017a) On the other hand, the major weight loss for PLA-functionalized GnP is observed between 200 - 400°C, corresponding to the decomposition of the polymer chains. In this temperature range, GNP Pyr-L showed a small weight loss ~ 3 wt.% (after subtraction of the weight loss corresponding to the pristine GnP), which corresponds to an adsorption concentration of 0.0535 mmol $g^{-1}$ or adsorption density of approx. 1400 chains $\mu m^{-2}$. The adsorption of the stereoisomers in the surface of GnP can be explained by i) the π-π interactions between pyrene moieties and GnP and ii) the intermolecular CH-π interactions between the –CH groups of PLLA and GnP,(Hu et al., 2009; Arenaza et al., 2015) the former assumed to be more significant. Therefore, the limited adsorption of Pyr-L is mainly ascribed to the low content of pyrene anchoring units, as each oligomeric PLA chain is end capped with a single pyrene group. This observation is consistent with previous studies on the non-covalent functionalization of carbon nanoparticles with pyrene containing polymers, where lower pyrene/polymer ratios showed a weakening of the polymer-carbon nanoparticle interactions.(Meuer et al., 2008; 2009)

GnP Pyr-L/Pyr-D SCs, prepared with the two different methodologies, were also investigated for their thermal decomposition. GnP Pyr-L/Pyr-D SCs exhibited a higher thermal stability, as observable from the onset temperature of 353.0 °C, compared to 322.3 °C for GnP Pyr-L to. Most interestingly, the amount of polymer chains adsorbed on the nanoplatelets is in both cases *ca.* 10 wt.%, (20 wt.% of the total initial amount of polymer) that corresponds to an adsorption concentration and density of 0.190 mmol g$^{-1}$ and approx. 12000 chains μm$^{-2}$, respectively. These values are one order of magnitude higher than those calculated for the GnP functionalized with only Pyr-L. The strong ability of carbon-based nanoparticles to induce the stereocomplex crystallization of PLA(Hu et al., 2009; Xu et al., 2010; Yang et al., 2016) together with the formation of SC precursors in the presence of DMF adsorbed on the GnP during the sonication process, favours the supramolecular functionalization of GnP with the SC.

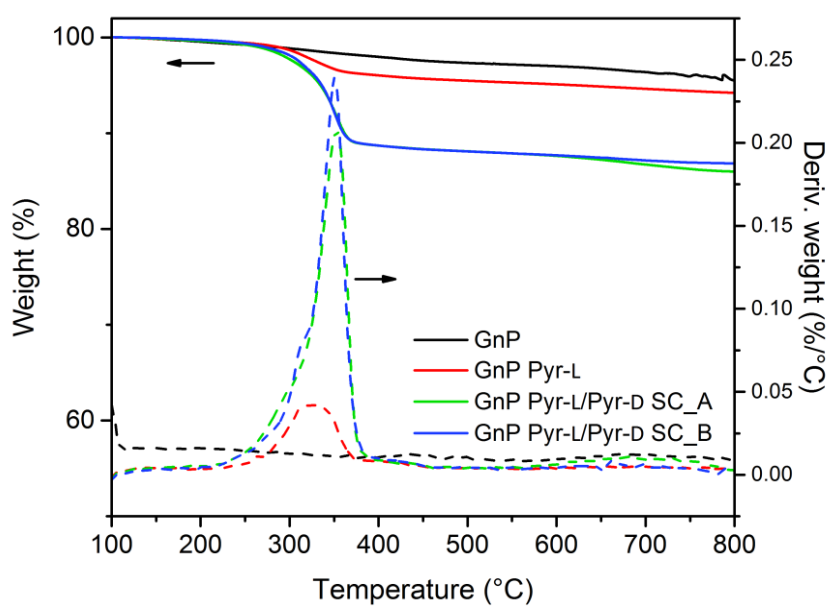

**Figure 5.** TGA curves of GnP and functionalized GnP.

DSC heating curves and XRD patterns of functionalized GnP (Figure 6) were further performed to discern the phenomena related to the adsorption of the stereoisomers on the GnP and the formation of the SC crystallites. The DSC thermogram of GnP Pyr-L did not show any clear melting peak of the homocrystals (Figure 6a) and no diffraction peaks of Pyr-L could be observed on the XRD pattern (Figure 6b). These results confirm the low amount of stereoisomer adsorbed on the basal plane of GnP, as already evidenced from TGA. On the other hand, GnP Pyr-L/Pyr-D SC exhibited clear evidences of PLA stereocomplexation, with significant differences as a function of the preparation method (Figure 6a inset). GnP Pyr-L/Pyr-D SC_A, prepared sonicating GnP in a mixed solution of the stereoisomers in DMF, showed a single $T_m$ at 222.8 °C, which is 15.3 °C higher than the $T_m$ of the SC in the absence of GnP, while showing the typical diffraction peaks (Figure 6b) of the SC crystallites at 2θ = 12.0, 20.8 and 24.1 °, assigned to the (110), (330)/(030) and (220) planes.(Tsuji, 2005; Bao et al., 2016) On the other hand, GnP Pyr-L/Pyr-D SC_B, where GnP was separately sonicated in the presence of one of the stereoisomers before mixing them to form the SC, exhibited two partially overlapped melting peaks at about 193 and 211°C (Figure 6a, inset), while displaying a XRD signal at the same position but with lower intensities, when compared to GnP Pyr-L/Pyr-D SC_A. The double melting peak has been previously observed in the precipitates of PDLA + PLLA mixtures of the stereoisomers in solution and explained by melting of less perfect crystallites and recrystallization into more stable crystallites.(Tsuji et al., 1992) It is worth noting that the main melting point of SCs in GnP Pyr-L/Pyr-D SC_B was found at a temperature (211.2 °C) that is significantly lower than the $T_m$ of GnP Pyr-L/Pyr-D SC_A. These differences highlight the role of both GnP and interactions of pyrene-terminated PLA oligomers to its surface. Indeed, in GnP Pyr-L/Pyr-D SC_B, the stereoisomers anchored onto the basal plane of GnP through pyrene terminals clearly affects the nucleation and growth of the SC crystallites. Hence, the probability of the intermolecular contact between the stereoisomers in solution and those on GnP increases, enhancing the stereocomplex crystallization from the surface of the nanoparticles.(Hu et al., 2009; Yang et al., 2016) However, the mobility of Pyr-L and Pyr-D chains adsorbed on the GnP is limited, organizing the SC crystallites in a poorly stable state, as observed by the DSC results. In GnP Pyr-L/Pyr-D SC_A, two competing effects are occurring simultaneously: the anchoring of the polymer chains to the nanoplatelets and the formation of the SC crystallites from enantiomers in solution. Here, the mobility of the polymer chains is promoted by sonication in solution together with the GnP, enhancing the stereocomplexation process, while the pyrene groups favour their adsorption on the basal planes of the GnP. As a direct consequence, more organized and stable crystallites are obtained in GnP Pyr-L/Pyr-D SC_A.

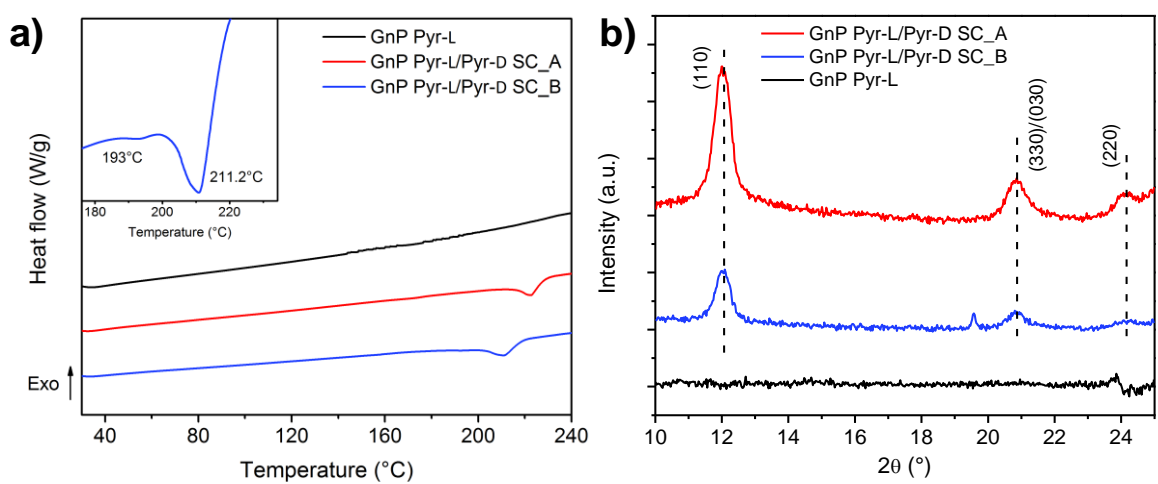

**Figure 6.** a) DSC thermograms of GnP Pyr-L and GnP Pyr-L/Pyr-D SC recorded during the second heating (10 °C min$^{-1}$). Inset shows the melting region of GnP Pyr-L/Pyr-D SC_B. b) Powder XRD patterns of GnP Pyr-L and GnP Pyr-L/Pyr-D SC.

Functionalized GnPs were used to prepare nanopapers, to assess the effect of the modification with Pyr-D/Pyr-L on the thermal transfer properties. In particular, the nanopapers have been prepared following the method described for GnP Pyr-L/Pyr-D SC_A, that has been previously confirmed that this strategy forms better SC crystallites. Furthermore, two nanopapers with increasing concentration of the stereoisomers were prepared to investigate the effect of the SC content on the properties of the nanopapers.

Nanopapers density was significantly decreased by the presence of Pyr-L/Pyr-D SC, in the range of 0.3÷0.4 g cm$^{-3}$, compared to 1.0 g cm$^{-3}$ for the pristine GNP nanopaper (Table 1). Despite the density is slightly decreasing with increasing the enantiomer concentration, the strong reduction in density is not consistent with the simple presence of the SC phase, based on the organic content, as determined by residue at 600°C in thermogravimetric analyses (Table 1 and Figure S6). Therefore, the nanopaper density appears to be mainly dependent on a different organization of the flakes, driven by the organic functionalization with Pyr-L/Pyr-D.

**Table 1.** Density ($\rho$) and in-plane ($\alpha_{n\parallel}$) and cross-plane ($\alpha_{n\perp}$) thermal diffusivities of nanopapers.

| Nanopaper | $\rho$ (g cm$^{-3}$) | SC content | $\alpha_\parallel$ (mm$^2$s$^{-1}$) | $\alpha_\perp$ (mm$^2$s$^{-1}$) |
|---|---|---|---|---|
| GnP | 1.03 ± 0.02 | - | 221.6 ± 4.2 | 0.96 ± 0.02 |
| GnP Pyr-L/Pyr-D SC_A | 0.40 ± 0.06 | 7.0 | 163.9 ± 3.6 | 0.53 ± 0.01 |
| GnP Pyr-L/Pyr-D SC_A′ | 0.35 ± 0.06 | 9.1 | 148.0 ± 2.3 | 2.85 ± 0.05 |
| GnP Pyr-L/Pyr-D SC_A″ | 0.32 ± 0.07 | 16.3 | 116.0 ± 15.8 | 4.50 ± 0.01 |

To investigate the microstructure of the nanopapers, FESEM analyses were carried out; top-view and cross-sectional images of the nanopapers are shown in Figure 7. The presence of the organic phase, corresponding to the Pyr-L/Pyr-D SC (Figures 7b, c and d), can be observed on the top-view images as thin coating onto the GnP flakes. The cross-sectional images of the nanopapers (Figure 7 insets) suggest the presence of the polymer chains, indeed modifies the structural organization of the GnP flakes. To further investigate the self-assembly of GnP flakes in the nanopaper, 2D-XRD measurements were carried out to qualitatively evaluate the degree of in-plane orientation. In Figure 8, 2D XRD patterns for the nanopapers of pristine GnP and GnP Pyr-L/Pyr-D SC are reported. As expected for nanopapers prepared by vacuum filtration from suspensions of 2D nanoparticles, a very high degree of in-plane preferential orientation was observed for pristine GnP. This is evidenced by the strongly variable intensity distribution of the 002 basal plane for graphite *vs*. the azimuthal angle, with the characteristic two maxima is opposite positions, confirming a very large fraction of the GnP flakes is oriented parallel to the plane of the nanopaper. In the presence of Pyr-L/Pyr-D SC, a slightly lower degree of orientation is generally observed, as suggested by a weaker orientation of the diffraction signal for 002 graphitic planes. This limited loss of orientation is particularly clear with in the cases of GnP Pyr-L/Pyr-D SC_A′ and GnP Pyr-L/Pyr-D SC_A″, suggesting that higher concentrations of SC have a role in the self-assembly of GnP nanoflakes during filtration. Furthermore, for GnP Pyr-L/Pyr-D inner diffraction circles are visible, corresponding to the XRD diffraction signals from the SC (Figure S7), with no significant preferential orientation (Figure 8c,d) and intensity proportional to the SC concentration determined by TGA.

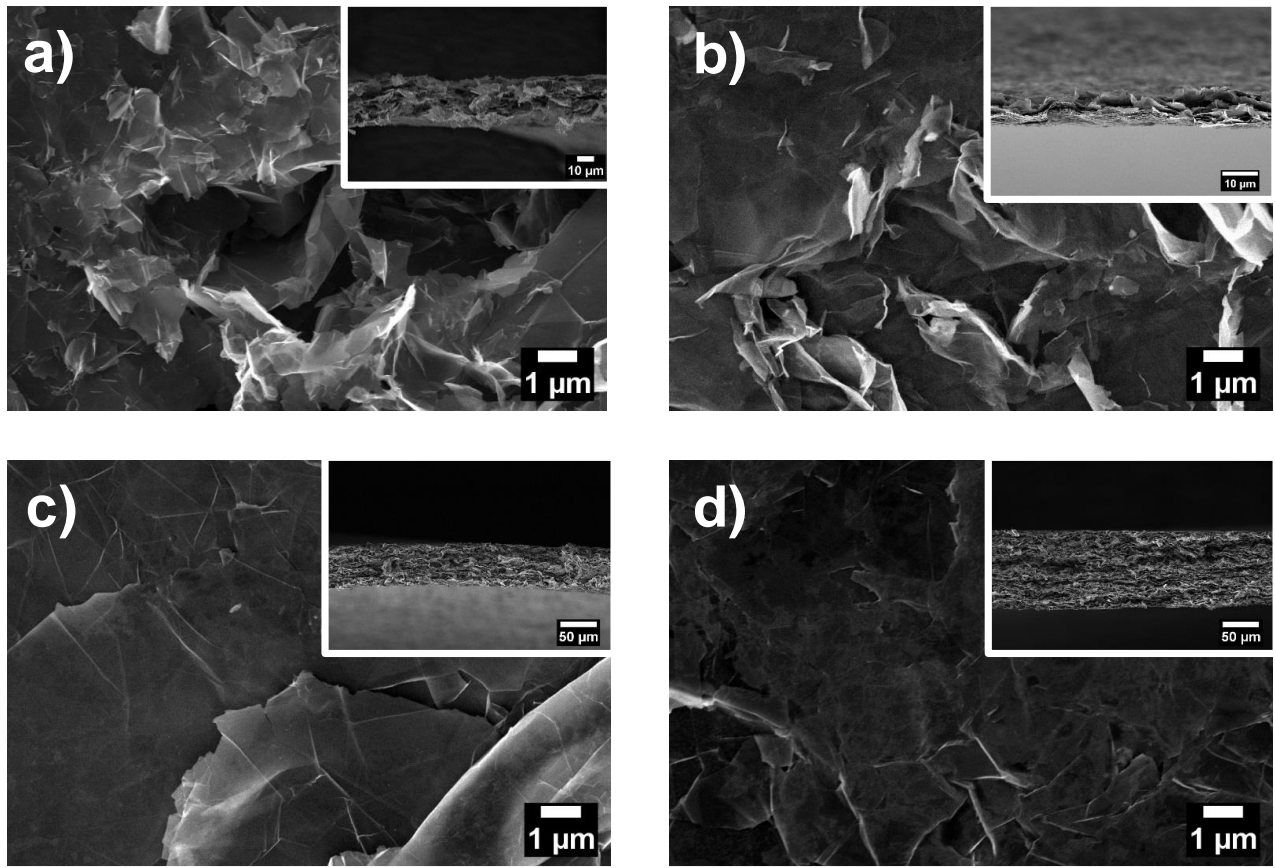

**Figure 7.** Top view FESEM images of GnP nanopapers: a) GnP, b) GnP Pyr-L/Pyr-D SC_A, c) GnP Pyr-L/Pyr-D SC_A′ and d) GnP Pyr-L/Pyr-D SC_A″. Insets are the corresponding cross-sectional FESEM images of the nanopapers.

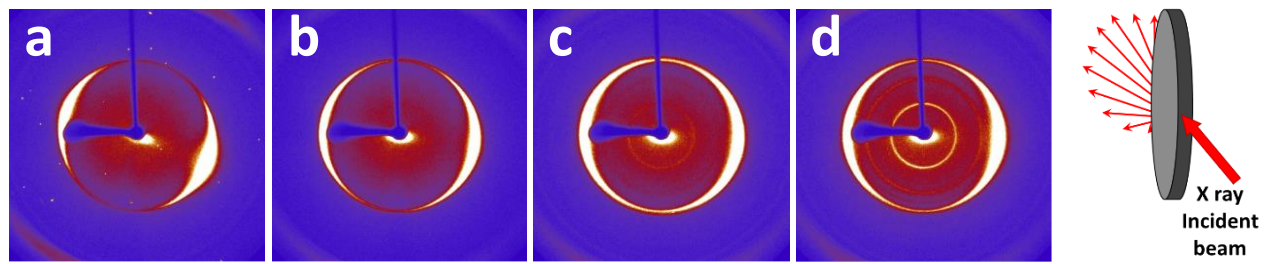

**Figure 8.** 2D XRD patterns measured via transmission geometry on nanopapers of pristine GnP (a) GnP Pyr-L/Pyr-D SC_A (b), GnP Pyr-L/Pyr-D SC_A′ (c) and GnP Pyr-L/Pyr-D SC_A″ (d). A schematic of pattern collection is also reported.

The thermal conductivities of the nanopapers were analyzed in both cross-plane and in-plane directions. Results reported in Table 1 and Figure 9 show a clear reduction of in-plane diffusivity in the presence of SC, which is expected, based on the low thermal conductivity of polymers. Furthermore, this reduction is a direct function of the fraction of SC in the nanopaper. Conversely, the cross plane diffusivity showed only a slight reduction at the lowest SC content, whereas at higher SC content the diffusivity is increased by a factor of 3 in GnP Pyr-L/Pyr-D SC_A′ and 4.5 in GnP Pyr-L/Pyr-D SC_A″, despite the increasing fraction of the polymeric phase.

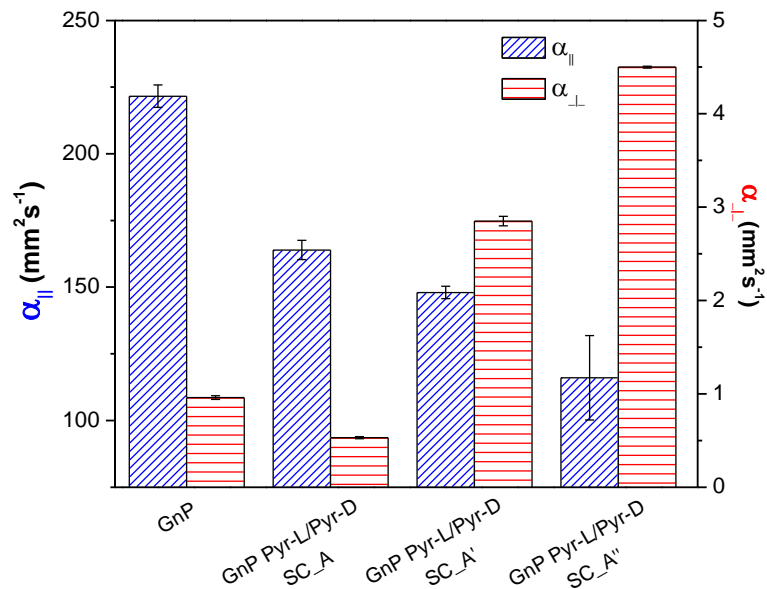

**Figure 9.** In-plane ($\alpha_{\parallel}$) and cross-plane ($\alpha_{\perp}$) thermal diffusivities for pristine and functionalized GnP nanopapers, with different contents of Pyr-L/Pyr-D SC.

The increase in cross-plane diffusivity may partially be related to a loss in nanoflakes orientation. As the fractions of flakes lying on direction tilted with respect to the nanopaper plane are expected to contribute more to the heat transfer across the nanopaper, owing to the well-known anisotropy of graphitic materials. However, in this case, the above commented differences in preferential orientation are not sufficient to explain the change in cross-plane diffusivity. Indeed, the lower degree of orientation was observed for GnP Pyr-L/Pyr-D SC_A, compared to pristine GnP, does not correspond to a better cross plane diffusivity. Furthermore, the significant differences between diffusivities for GnP Pyr-L/Pyr-D SC_A′ and GnP Pyr-L/Pyr-D SC_A″ are not reflected in terms of preferential orientations, proven to be comparable. Therefore, the enhanced efficiency on cross-plane heat transfer appears to be related to the organization of the SC phase, acting as thermal bridge between GnP flakes. In fact, crystallinity and orientation of the macromolecules are well-known to dramatically affect the thermal conductivity of the polymeric materials (Choy et al., 1980; Choy et al., 1993; Singh et al., 2014) and the organization of a thermally stable and highly crystalline SC phase, strongly interacting to the surface of GnP via the pyrene terminal groups, appears to have an effective role in promoting heat transfer between overlapped GnP flakes.

**Conclusions**

Non-covalent functionalization of graphite nanoplatelets (GnP) was obtained with *ad hoc* synthesized pyrene chain-ended oligomers of stereoregular polylactic acid (Pyr-L and Pyr-D), exploiting the π- π stacking interactions with GnP surface. Stereocomplexation of Pyr-L and Pyr-D onto GnP was obtained both *via* one-pot preparation from a mixture of GnP/Pyr-L/Pyr-D, as well as by mixing of GnP/Pyr-L and GnP/Pyr-D suspensions, with the former procedure allowing the preparation of better organized SC domains. Nanopapers were prepared at different SC fraction, evidencing the dominant role of the polymeric functionalization in the self-assembly of the GnP flakes. In particular, lower densities and reduced orientations of the nanopapers were induced by the presence of SC, acting as a polymeric cross-linkers between different GnP moieies. Most interestingly, the presence of PLA SC between GnP flakes was found to enhance cross-plane heat transfer in GnP Pyr-L/Pyr-D SC nanopapers, explained in terms of the contribution of local crystallinity to the reduction of thermal resistance at the interface between GnP flakes. Therefore, the possibility to control the thermal conductivity anisotropy of GnP nanostructures, by the self-assembly of organic polymeric functionalization, was demontrated. The proposed approach can be generally applied to the physical properties modification of engineered nanostructures based on graphene and graphene related materials.

**Acknowledgements**

This work has received funding from the European Research Council (ERC) under the European Union's Horizon 2020 research and innovation programme grant agreement 639495 — INTHERM — ERC-2014-

The authors gratefully acknowledge dr. Julio Gomez (Avanzare Innovación Tecnólogica, E) for providing GnP, dr. Mauro Raimondo (Politecnico di Torino, I) for FESEM analyses, dr. Salvatore

©2019. This manuscript version is made available under the CC-BY-NC-ND 4.0 license
http://creativecommons.org/licenses/by-nc-nd/4.0/

Marubayashi, H., Asai, S., and Sumita, M. (2012). Complex Crystal Formation of Poly(l-lactide) with Solvent Molecules. *Macromolecules* 45(3), 1384-1397. doi: 10.1021/ma202324g.

Marubayashi, H., Asai, S., and Sumita, M. (2013). Guest-Induced Crystal-to-Crystal Transitions of Poly(l-lactide) Complexes. *The Journal of Physical Chemistry B* 117(1), 385-397. doi: 10.1021/jp308999t.

Meng, Y., Wang, K., Zhang, Y., and Wei, Z. (2013). Hierarchical Porous Graphene/Polyaniline Composite Film with Superior Rate Performance for Flexible Supercapacitors. *Advanced Materials* 25(48), 6985-6990. doi: doi:10.1002/adma.201303529.

Meuer, S., Braun, L., and Zentel, R. (2008). Solubilisation of multi walled carbon nanotubes by [small alpha]-pyrene functionalised PMMA and their liquid crystalline self-organisation. *Chemical Communications* (27), 3166-3168. doi: 10.1039/B803099E.

Meuer, S., Braun, L., and Zentel, R. (2009). Pyrene Containing Polymers for the Non-Covalent Functionalization of Carbon Nanotubes. *Macromolecular Chemistry and Physics* 210(18), 1528-1535. doi: doi:10.1002/macp.200900125.

Michalski, A., Socka, M., Brzezinski, M., and Biela, T. (2018). Reversible Supramolecular Polylactides Gels Obtained via Stereocomplexation. *Macromolecular Chemistry and Physics* 219(9). doi: ARTN 1700607

10.1002/macp.201700607.

Mittal, V. (2014). Functional Polymer Nanocomposites with Graphene: A Review. *Macromolecular Materials and Engineering* 299(8), 906-931. doi: doi:10.1002/mame.201300394.

Papageorgiou, D.G., Kinloch, I.A., and Young, R.J. (2017). Mechanical properties of graphene and graphene-based nanocomposites. *Progress in Materials Science* 90, 75-127. doi: https://doi.org/10.1016/j.pmatsci.2017.07.004.

Parviz, D., Das, S., Ahmed, H.S.T., Irin, F., Bhattacharia, S., and Green, M.J. (2012). Dispersions of Non-Covalently Functionalized Graphene with Minimal Stabilizer. *ACS Nano* 6(10), 8857-8867. doi: 10.1021/nn302784m.

Potts, J.R., Dreyer, D.R., Bielawski, C.W., and Ruoff, R.S. (2011). Graphene-based polymer nanocomposites. *Polymer* 52(1), 5-25. doi: https://doi.org/10.1016/j.polymer.2010.11.042.

Renteria, J.D., Ramirez, S., Malekpour, H., Alonso, B., Centeno, A., Zurutuza, A., et al. (2015). Strongly Anisotropic Thermal Conductivity of Free-Standing Reduced Graphene Oxide Films Annealed at High Temperature. *Advanced Functional Materials* 25(29), 4664-4672. doi: doi:10.1002/adfm.201501429.

Rodriguez-Perez, L., Herranz, M.a.A., and Martin, N. (2013). The chemistry of pristine graphene. *Chemical Communications* 49(36), 3721-3735. doi: 10.1039/C3CC38950B.

# Stereocomplexation of Poly(Lactic Acid)s on Graphite Nanoplatelets: from Functionalized Nanoparticles to Self-Assembled Nanostructures.

Matteo Eleuteri, Mar Bernal, Marco Milanesio, Orietta Monticelli, Alberto Fina

**1.1 Characterization of Pyr-L**

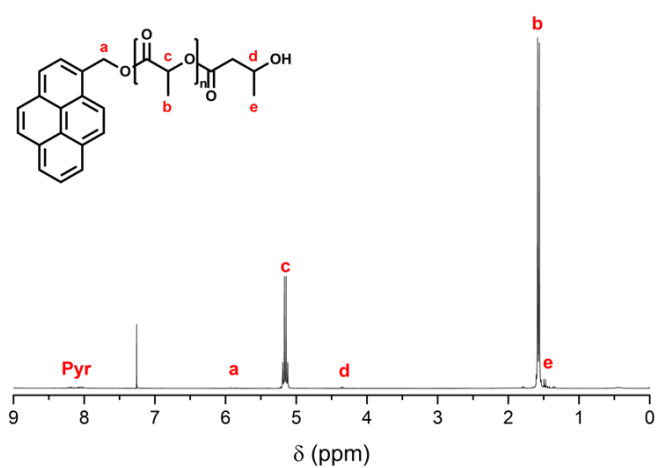

**Figure S1.** $^1$H NMR spectrum in CDCl$_3$ of Pyr-L.

The UV-Vis spectra of Pyr-L in DMF at a concentration of 0.5 mg mL$^{-1}$ is shown in Figure S2. The typical absorbance bands of pyrene in the region of 270 – 350 nm can be clearly observed.

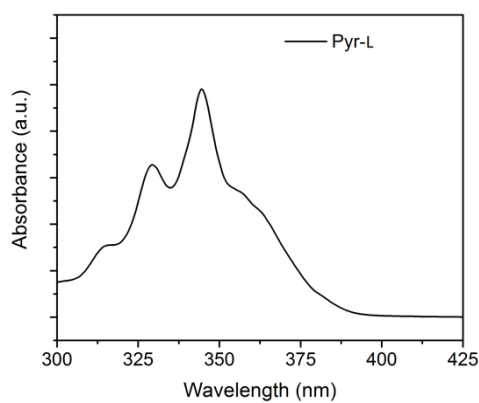

**Figure S2.** UV-Vis absorption spectra of Pyr-L in DMF at a concentration of 0.5 mg mL$^{-1}$.

## 1.2 Effect of the Solvent on the Pyr-L/Pyr-D Stereocomplexation

Pyr-L/Pyr-D sterecomplexes were prepared from solutions in different solvents, namely chloroform and N,N-dimethylformamide. The first is a conventional solvent for polylactic acid, whereas the second was selected for its well-known affinity to GnP. While stereocomplexation is observed to occur in both solvents, results from thermal analysed as well as spectroscopy evidence for dramatic differences in organization of SC crystallites, as a function of the PLA oligomers with the solvent.

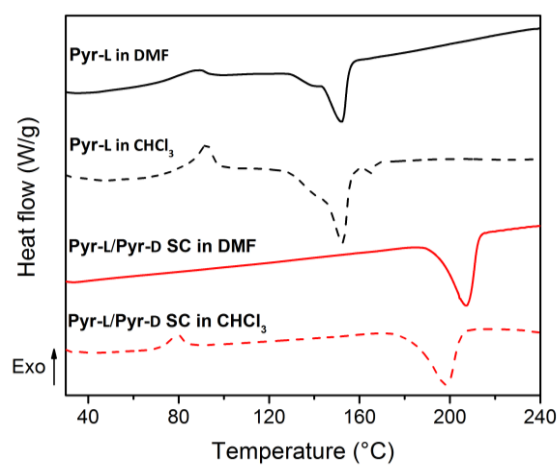

**Figure S3.** DSC thermograms of Pyr-L and Pyr-L/Pyr-D SC, recrystallized from DMF and CHCl$_3$, recorded during the second heating (10 °C min$^{-1}$).

**Table S1.** Thermal parameters of Pyr-L and Pyr-L/Pyr-D SC obtained after crystallization in DMF and CHCl$_3$: cold crystallization temperature ($T_{cc}$) and enthalpy ($\Delta H_{cc}$), melting temperature ($T_m$) and enthalpy ($\Delta H_m$) and temperature corresponding to the maximum DTG peaks ($T_{max}$) obtained from the first derivative curve of the TGA thermograph of Pyr-L and Pyr-L/Pyr-D SC.

| Sample | $T_{cc}^a$ (°C) | $\Delta H_{cc}^a$ (J g$^{-1}$) | $T_m^a$ (°C) | $\Delta H_m^a$ (J g$^{-1}$) | $T_{max}^b$ (°C) |
|---|---|---|---|---|---|
| **Pyr-L (DMF)** | 87.1 | 7.3 | 138.5, 152.2 | 39.3 | 264.2 |
| **Pyr-L (CHCl$_3$)** | 92.2 | 11.0 | 140.6, 152.5, 165.1 | 38.8 | 274.0 |
| **Pyr-L/Pyr-D SC (DMF)** | - | - | 207.5 | 45.1 | 296.6 |
| **Pyr-L/Pyr-D SC (CHCl$_3$)** | 79.3 | 5.6 | 198.7 | 43.5 | 293.4 |

[a] Determined by DSC analysis, second heating (10 °C min$^{-1}$).
[b] Determined by TGA analysis.

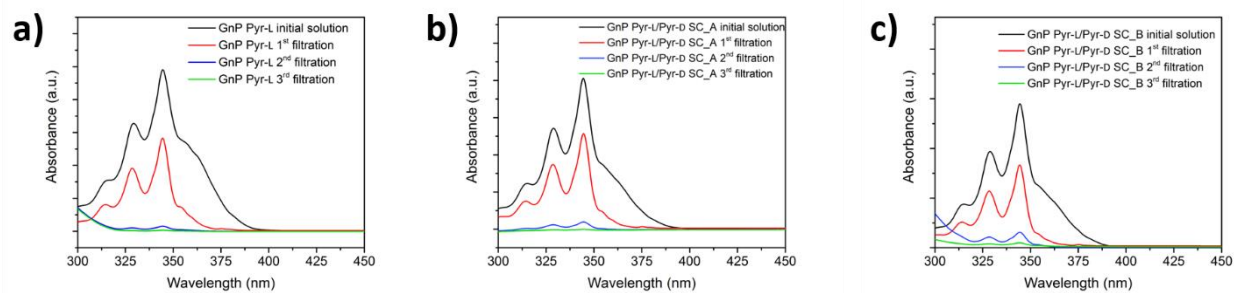

**Figure S4.** UV-Vis spectra obtained after dispersion-filtration cycles during the preparation of a) GnP Pyr-L, b) GnP Pyr-L/Pyr-D SC_A and c) GnP Pyr-L/Pyr-D SC_B.

### 1.3 XPS analysis

Pyr-L and Pyr-L/Pyr-D SC were first measured as reference samples in order to facilitate the interpretation of the XPS spectra of supramolecularly functionalized GnP (Figure S5). In the case of Pyr-L and Pyr-L/Pyr-D SC, the $C_{1s}$ spectra show the components related to the polymers structure:(Shakesheff et al., 1997; Jordá-Vilaplana et al., 2014) the peak at ~ 284.9 eV attributed to aliphatic carbons (-$CH_2$, -$CH_3$), the band at 285.5 eV (C-OH), the C-O-C peak at 286.8 eV and the 288.9 eV binding energy of the -C=O. The deconvolution of the $C_{1s}$ spectra of the functionalized GnP are dominated by the peak at 284.4 eV, related to the $sp^2$ C-C bonds and thus the differentiation of the aliphatic carbons of the polymer backbone remains difficult, suggesting for a limited fraction of PyrL or PyrD onto GnP. The $O_{1s}$ spectra (Figure 5b) of Pyr-L and Pyr-L/Pyr-D SC were deconvoluted into the three bands typical of the composition of PLA: the O-C=O at 531.5 eV, the C-O-C at 533 eV and the C-OH at 534.2 eV. On the other hand, the $O_{1s}$ spectrum of GnP showed only two components due to the C=O at 530.5 eV and a broad band related to the C-O-C group. The successful functionalization of the GnP with the polymer chains is confirmed by the presence on the $O_{1s}$ spectra of the corresponding oxygen bands observed in the $O_{1s}$ spectra of Pyr-L and Pyr-L/Pyr-D SC. The values of the at.% of each oxygen functional group are summarized in Table S2. Atomic ratios (at.%) were calculated from experimental intensity ratios and normalized by atomic sensitivity factors. The $C_{1s}$ and $O_{1s}$ peaks were fitted as following the previously described procedure.(Bernal et al.)

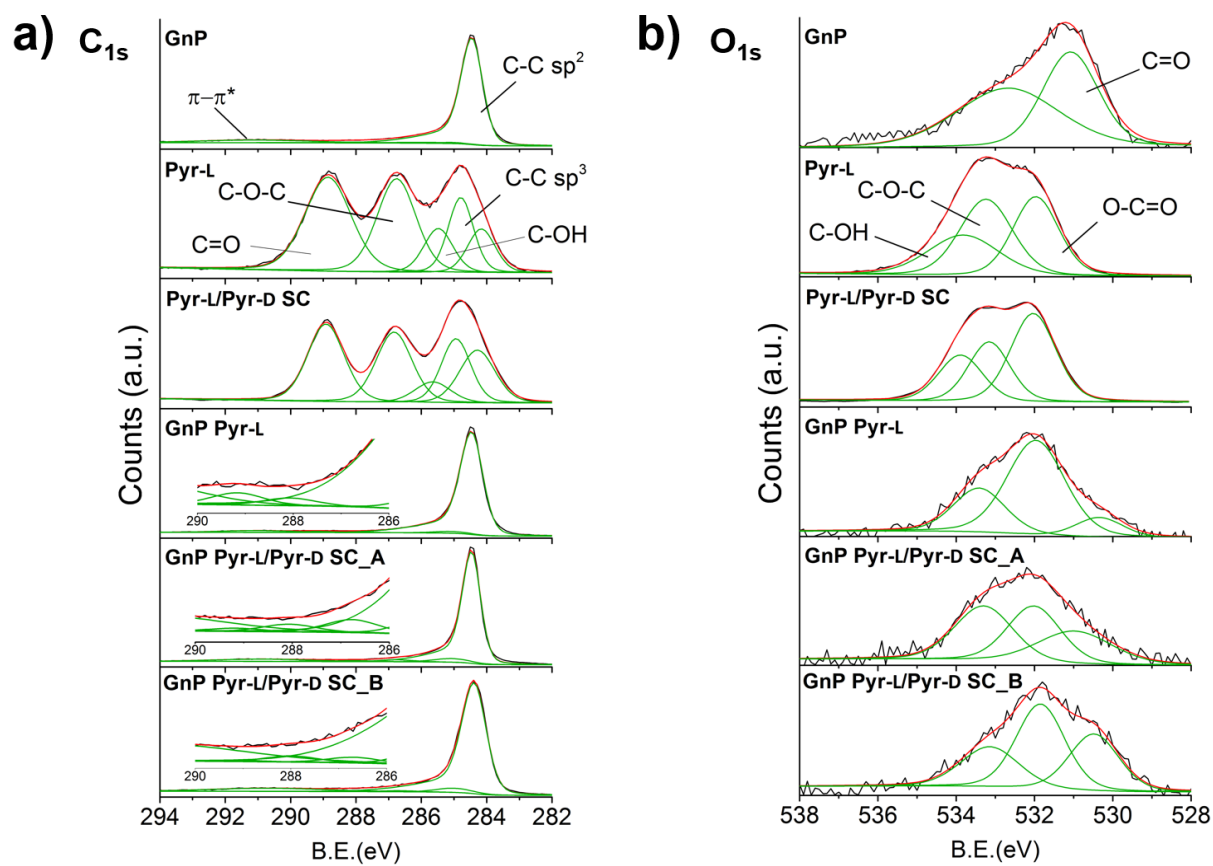

**Figure S5.** $C_{1s}$ and $O_{1s}$ XPS spectra of GnP, Pyr-L, Pyr-L/Pyr-D SC_A, Pyr-L/Pyr-D SC_B, GnP Pyr-L, GnP Pyr-L/Pyr-D SC_A and GnP Pyr-L/Pyr-D SC_B.

**Table S2.** XPS peak assignment and atomic percentage (at.%) for Pyr-L, Pyr-L/Pyr-D SC, GnP, GnP Pyr-L, GnP Pyr-L/Pyr-D SC _A and GnP Pyr-L/Pyr-D SC _B.

| Sample | Total $C_{1s}$ at.% | Total $O_{1s}$ at.% | Contributions from O to $C_{1s}$ spectra at.% | | |
|---|---|---|---|---|---|
| | | | C-OH | C-O-C | C=O |
| **Pyr-L** | 59.2 | 40.8 | 6.1 | 26.1 | 20.7 |
| **Pyr-L/Pyr-D SC** | 59.2 | 40.8 | 4.5 | 22.7 | 17.0 |
| **GnP** | 96.5 | 3.5 | 0.9 | 0.8 | 1.8 |
| **GnP Pyr-L** | 96.4 | 3.6 | 0.9 | 1.1 | 1.5 |
| **GnP Pyr-L/Pyr-D SC _A** | 95.5 | 4.5 | 0.6 | 3.3 | 1.2 |
| GnP Pyr-L/Pyr-D SC _B | 96.0 | 4.0 | 0.8 | 1.5 | 1.8 |

## 1.4 Adsorption concentration and adsorption density

The adsorption concentration and density of the stereoisomers and the SCs were determined from the TGA data of the fuctionalized GnP.

The adsorption concentration (mmol g$^{-1}$) was calculated from the mmol of polymer chains adsorbed per gram of framework graphene carbon according to the literature:(Hu et al., 2017)

$$GC\left(\frac{mmol}{g}\right) = \frac{\left(\frac{M_p/(M_p + M_G)}{M_{wp}}\right)}{\left(1 - \frac{M_p}{(M_p + M_G)}\right)} \times 1000$$

where $M_p$ is the total mass of adsorbed polymer, $M_G$ is the total mass of GnP and $M_{wp}$ is the molecular weight of the polymer in g mol$^{-1}$.

The adsorption density, defined by the number of polymers per area of GnP surface, *N/A*, was calculated as:(Chadwick et al., 2013)

$$\frac{N}{A} = \frac{M}{A}\frac{M_p}{M_G}\frac{1}{M_{wp}}$$

where *M/A* is the mass per unit area of graphene estimated as 7.48 x 10$^{-7}$ kg m$^{-2}$ (assuming an average area of GnP of 100 μm$^2$ and a density of 2200 kg m$^{-3}$) and $M_{wp}$ is converted to kg molecule$^{-1}$.

## 1.5 Additional characterization of GNP Pyr-L/Pyr-D SC nanopapers

Thermogravimetry was used to assess the content of organic fraction, namely the Pyr-l/Pyr-D SC in the nanopapers prepared form different enantiomers concentrations. Results are reported in Figure S6.

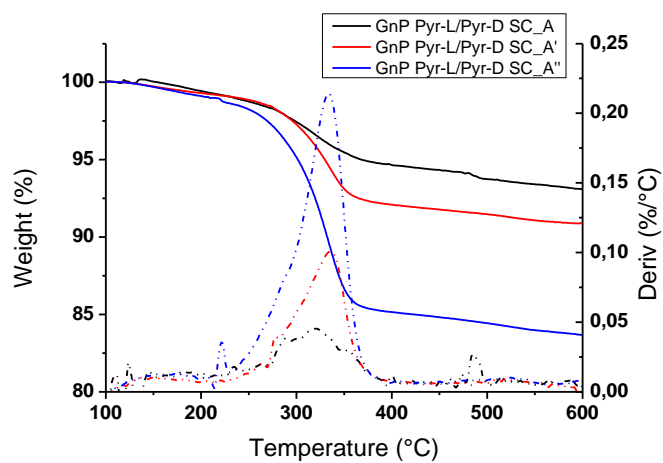

**Figure S6.** Thermogravimetric plots for nanopapers of GnP Pyr-L/Pyr-D SC prepared with different enantiomers concentrations

2D XRD were reduced to intensity vs. 2θ profiles, to investigate preferential orientations in the nanopapers and well as to obtain averaged XRD profiles over the whole possible orientations. Results are reported in Figure S7.

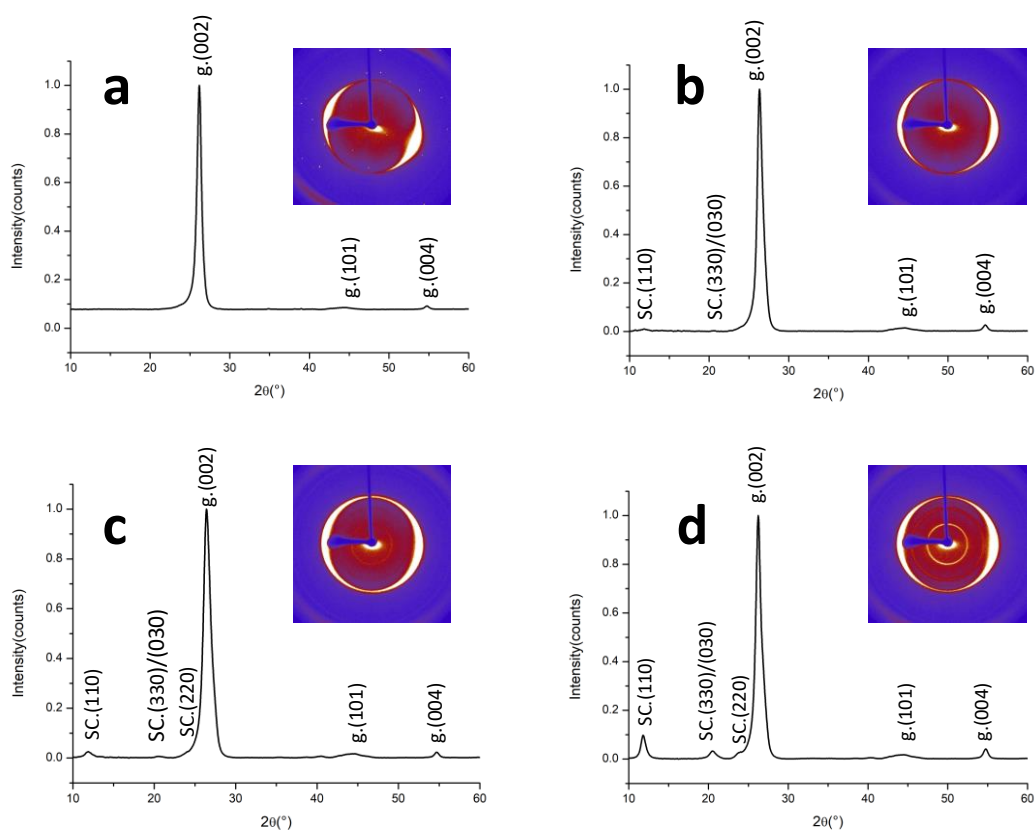

**Figure S7.** XRD spectra from 0-360° intergration of 2D Patterns obtained on nanopapers of pristine GnP (a) GnP Pyr-L/Pyr-D SC_A (b), GnP Pyr-L/Pyr-D SC_A′ (c) and GnP Pyr-L/Pyr-D SC_A″ (d). Peaks are identified with their crystalline plane s assignment graphene (g) or stereocomplex (SC)